\documentclass[aps,pre,twocolumn,superscriptaddress,nobibnotes,nodoi,noeprint]{revtex4-2}

\usepackage{graphicx}
\usepackage{amsmath,amssymb,physics,bm,float}
\usepackage{soul}
\usepackage[colorlinks=true,citecolor=blue,urlcolor=blue,linkcolor=blue]{hyperref}
\usepackage{natbib}
\usepackage{upgreek}
\usepackage{mathtools}
\usepackage{dcolumn}

\newcommand{\nb}{n_{\rm b}}
\newcommand{\mb}{m_{\rm b}}

\newcommand{\nc}{n_{\rm c}}

\newcommand{\nbc}{(n_{\rm c}+n_{\rm b})}

\newcommand{\Nsol}{N_{\rm sol}}
\newcommand{\nsol}{n_{\rm sol}}
\newcommand{\msol}{m_{\rm sol}}
\newcommand{\Nb}{N_{\rm b}}
\newcommand{\Npre}{N_{\rm pre}}

\newcommand{\lambdasol}{\lambda_{\rm sol}}
\newcommand{\Displ}{\mathcal{D}}
\newcommand{\tausol}{\tau_{\rm sol}}

\DeclareMathOperator*{\ourmod}{\,mod\,}

\begin{document}

\title{Cluster sizes, particle displacements and currents in transport\\ mediated by solitary cluster waves}

\author{Alexander P. Antonov}
\email{alantonov@uos.de}
\affiliation{Universit{\"a}t Osnabr{\"u}ck,
Fachbereich Mathematik/Informatik/Physik,
Institut f{\"u}r Physik,
Barbarastra{\ss}e 7,
D-49076 Osnabr{\"u}ck,
Germany}
\affiliation{Institut f{\"u}r Theoretische Physik II: Weiche Materie, 
Heinrich-Heine-Universit{\"a}t D{\"u}sseldorf, 
Universit\"atsstra{\ss}e 1,
D-40225 D{\"u}sseldorf, 
Germany}

\author{Annika Vonhusen}
\email{avonhusen@uos.de}
\affiliation{Universit{\"a}t Osnabr{\"u}ck, 
Fachbereich Mathematik/Informatik/Physik,
Institut f{\"u}r Physik,
Barbarastra{\ss}e 7,
D-49076 Osnabr{\"u}ck,
Germany}

\author{Artem Ryabov}
\email{rjabov.a@gmail.com}
\affiliation{Charles University, 
Faculty of Mathematics and Physics, 
Department of Macromolecular Physics, 
V Hole\v{s}ovi\v{c}k\'ach 2, 
CZ-18000 Praha 8, 
Czech Republic}

\author{Philipp Maass}
\email{maass@uos.de}
\affiliation{Universit{\"a}t Osnabr{\"u}ck,
Fachbereich Mathematik/Informatik/Physik,
Institut f{\"u}r Physik,
Barbarastra{\ss}e 7,
D-49076 Osnabr{\"u}ck,
Germany}

\date{\today}

\begin{abstract}
In overdamped particle motion across periodic landscapes, solitary
cluster waves can occur at high particle densities and lead to
particle transport even in the absence of thermal noise.  Here we show
that for driven motion under a constant drag, the sum of all particle
displacements per soliton equals one wavelength of the periodic
potential.  This unit displacement law is used to determine particle
currents mediated by the solitons.  We furthermore derive properties
of clusters involved in the wave propagation as well as relations
between cluster sizes and soliton numbers.
\end{abstract}

\maketitle

\section{Introduction}
\label{sec:Introduction}

Collective transport of particles in crowded systems frequently
involves coherent motion of particle assemblies.  Important examples
are driven dynamics in biological pores and synthetic channels
\cite{Karnik/etal:2007, Misiunas/Keyser:2019, Su/etal:2022},
intracellular particle transport \cite{Bressloff/Newby:2013},
crowdion-mediated atomic diffusion on solid surfaces
\cite{Xiao/etal:2003}, and driven motions of micron-sized particles in
colloidal suspensions \cite{Korda/etal:2002, Bohlein/etal:2012,
  Bohlein/Bechinger:2012, Bohlein/Bechinger:2012, Tierno/Fischer:2014,
  Tierno/etal:2014, Juniper/etal:2015, Cao/etal:2019b,
  Stoop/etal:2020, Mirzaee-Kakhki/etal:2020, Lips/etal:2021,
  Leyva/etal:2022}.

A minimal model for studying such cluster-mediated collective dynamics
is that of hardcore interacting particles performing overdamped
Brownian in periodic potentials \cite{Lips/etal:2019,
  Castaneda-Priego/etal:2025}.  This model has been termed Brownian
asymmetric simple exclusion process (BASEP) \cite{Lips/etal:2018}
since it resembles the prominent asymmetric simple exclusion process,
a fundamental model in nonequilibrium statistical mechanics
\cite{Derrida:1998, Schuetz:2001, Schmittmann/Zia:1998, Mallick:2015}.

Recently, solitary cluster waves were theoretically predicted
\cite{Antonov/etal:2022a} to occur in the one-dimensional BASEP and
experimentally confirmed for driven Brownian motion of microparticles
across a periodic potential \cite{Cereceda-Lopez/etal:2023}.  The
solitary cluster waves emerge at high particle densities, when the
filling of potential wells exceeds a certain limit. There exists a
maximal filling, where the particles arrange into mechanically stable
configurations in the periodic landscape under a constant drag
force. The ensemble of these configurations defines the presoliton
state. For higher fillings, running states are formed with particle
transport mediated by cluster waves.  The waves propagate by
attachment and detachment events, where one cluster gets in contact
with or looses contact from another cluster.

Solitary cluster wave dynamics can be synchronized with an external
oscillatory driving \cite{Mishra/etal:2025}, leading to phase-locked
values of soliton-mediated particle currents and Shapiro steps in the
change of currents upon modifying the period-averaged driving
force. When analyzing properties of the phase-locked currents, a unit
displacement law (UDL) was conjectured, saying that the sum of all
particle displacements during one period of soliton motion is equal to
the wavelength of the period potential.

Here we prove the UDL for the sinusoidal potential and weak constant
drag force.  The proof suggests that the UDL is valid also for large
drag forces and arbitrary periodic potentials and we test this by
numerical simulations.  Based on the UDL we derive cluster sizes
involved in the soliton propagation, relations to the number of
solitons occurring in the system, and particle currents mediated by
solitary cluster waves.

Considering the complexity of presoliton states, the UDL is
surprising.  In a presoliton state, particles can arrange into
different configurations, where in each configuration a different set
of cluster sizes is present.  By contrast, particle configurations are
made of same cluster types in a running state.  This aspect of
heterogeneity of presoliton states and homogeneity of soliton-carrying
states was not addressed in the previous work
\cite{Antonov/etal:2024}. We discuss it in this study and show that
the heterogeneity in presoliton states is decreasing with driving
force and system size.  In the thermodynamic limit, only the number of
the largest stabilizable clusters is extensive.

The UDL is remarkable also in view of the fact that individual
particle displacements in one soliton period are different.

\begin{figure*}[t!]
\centering
\includegraphics[width=\textwidth]{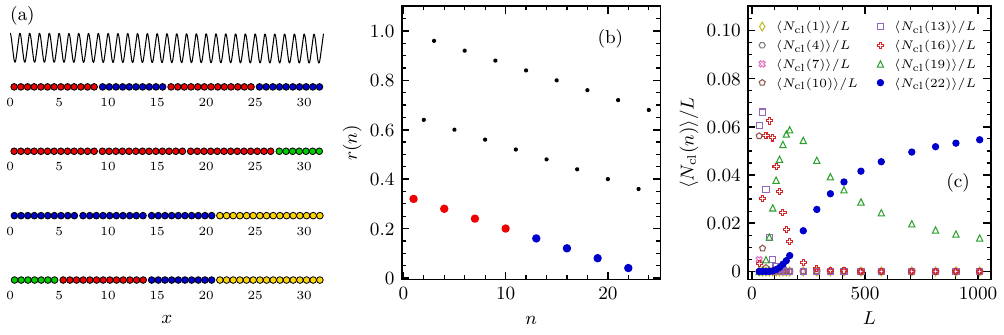}
\caption{Particle clusters in the presoliton state for particle
  diameter $\sigma=0.68$ and drag force $f=10^{-3}$, representing the
regime of infinitesimal $f$. (a)
  Examples of four simulated particle configurations in the presoliton
  state of a system with $L=32$ potential wells, obtained by evolving initial configurations with randomly
  chosen particle positions until a mechanically stable state is
  reached. In these configurations, stable clusters of sizes $n=7$ (green), 10 (blue), 13 (red), and 16 (yellow) appear.
  (b) Residual free space $r(n)$ [Eq.~\eqref{eq:r(n)}] as a
  function of $n$. Bold circles give $r(n)$ of the possible
  mechanically stable $n$-clusters according to the free space
  theorem. These $n$-clusters are stable and can occur in
  configurations of the presoliton state for $f=0^+$. For finite
  $f=0.05$, the red circles correspond to possible mechanically stable
  $n$-clusters, while the blue ones to $n$-clusters that become
  unstable due to the larger $f$. (c) Mean cluster numbers
  $\langle N_{\rm cl}(n)\rangle/L$ per potential well as a function of
  $L$, showing that only $\langle N_{\rm cl}(\nb=22)\rangle$ is
  extensive in system size. The averaging $\langle\ldots\rangle$ was
  performed over 500 initial particle configurations.}
\label{fig:presoliton_state_f=0}
\end{figure*}

\section{Driven overdamped motion of hard spheres in periodic potentials}
\label{sec:model}

We consider $N$ hard spheres that are dragged by a constant force $f$
accross a one-dimensional periodic potential $U(x)$ with wavelength
$\lambda$ in a fluidic environment. If the potential barriers are much
larger than the thermal energy, thermal noise effects become
negligible and equation of motions can be written as
\cite{Antonov/etal:2024}
\begin{equation}
\frac{\dd x_i}{\dd t}=f - \frac{\dd U}{\dd x_i}\,,\hspace{1em}i=1,\ldots,N\,,
\label{eq:langevin-zero-noise}
\end{equation}
where $x_i$ are the particle positions. They satisfy the hard-sphere
constraints
\begin{equation}
|x_j-x_i| \ge \sigma\,,
\label{eq:hard-sphere-constraints}
\end{equation}
where $\sigma$ is the particle diameter.
Equations~\eqref{eq:langevin-zero-noise} are given in dimensionless
units. We have taken the wavelength $\lambda$ and barrier height $U_0$
of the periodic potential $U(x)$ as length and energy unit, and
$\lambda^2/\mu U_0$ as the time unit, where $\mu$ is the particle
mobility or inverse friction coefficient.  The system size $L$ is an
integer, $L\in\mathbb{N}$, and periodic boundary conditions are used.

For the sinusoidal potential,
\begin{equation}
U(x) = \frac{1}{2}\cos(2\pi x)\,.
\label{eq:cosine-potential}
\end{equation}
As an example for a non-sinusoidal potential, we choose the smoothed
triangle wave
\begin{equation}
U_\delta(x)=\frac{1}{2}\left(1-\frac{2}{\pi}\arccos[(1-\delta)\cos(2\pi x)]\right)\,,
\label{eq:triangle-wave-potential}
\end{equation}
where for $\delta>0$ the cusps of the triangles are rounded.

We consider particle diameters $\sigma<1$ in the following. Dynamics
for larger particle diameters $\sigma+j$, $j\in\mathbb{N}$ can be
mapped onto that for $\sigma\in[0,1)$, because the equations of motion
  \eqref{eq:langevin-zero-noise} are invariant under a transformation
  $x_i\to x_i-j$; for details, see \cite{Lips/etal:2019,
    Antonov/etal:2024}.

To solve the equations of motions \eqref{eq:langevin-zero-noise}, we
apply the Brownian cluster dynamics method \cite{Antonov/etal:2025}.

\section{Heterogeneous presoliton and unique soliton-carrying states}
\label{sec:presoliton-running-state}
In a recent study \cite{Antonov/etal:2024}, a detailed theory was
presented for soliton formation, propagation and soliton-mediated
particle currents in the tilted sinusoidal potential. An essential
concept in this theory is the presoliton state.  It is the
mechanically stable state with largest number of particles.

The presoliton state can be realized by different particle
configurations.  Examples are shown in
Fig.~\ref{fig:presoliton_state_f=0}(a) for the sinusoidal potential
with $L=32$ wells, $\Npre=45$ particles, and particle diameter
$\sigma=0.68$. The drag force is $f=10^{-3}$, which represents the
regime of infinitesimal drag force $f=0^+$.  We generated the
mechanically stable configurations by evolving randomly chosen initial
particle positions.  In each configuration, clusters appear, where $n$
particles are in contact.  Specifically, we observe $n$-clusters with
$n=7$, 10, 13, and 16 in the configurations in
Fig.~\ref{fig:presoliton_state_f=0}(a). Which cluster sizes occur is
determined by the initial conditions.

Possible mechanically stable clusters in particle configurations of a
presoliton state can be determined from a free space theorem
\cite{Antonov/etal:2024}. It states:\\
\hspace*{0.1\columnwidth}\parbox[t]{0.8\columnwidth}{\it 
An  $n$-cluster is stabilizable, if and only if its residual free space is 
smaller than that of $n'$-clusters with $n'<n$.}

\begin{figure*}[t!]
\centering
\includegraphics[width=\textwidth]{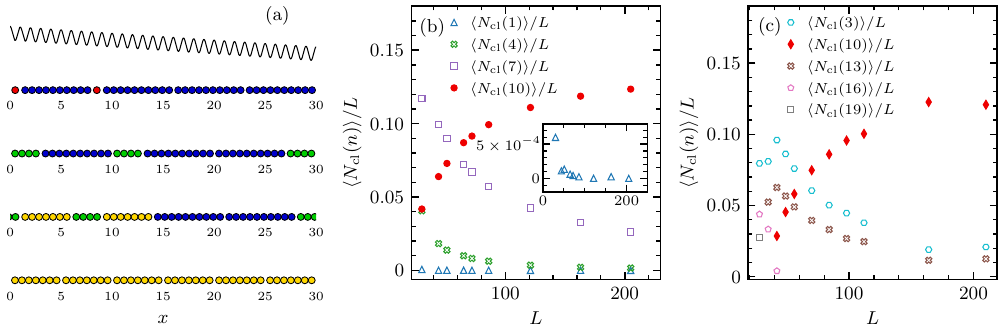}
\caption{Particle clusters in the presoliton and soliton-carrying running states for the same
particle diameter $\sigma=0.68$ as in Fig.~\ref{fig:presoliton_state_f=0} but for finite drag force $f=0.05$. 
(a) Examples of four simulated particle configurations in the presoliton state of a system with $L=32$ potential wells.
Stable clusters of sizes $n=1$ (red), $n=4$ (green), 7 (yellow), and 10 (blue) appear.
(b)~Mean cluster numbers $\langle N_{\rm cl}(n)\rangle/L$ per potential well as a function of $L$.
Only $\langle  N_{\rm cl}(n)\rangle$ for $n=\nb=10$ is extensive in system size.
The inset shows a zoom-in of $\langle N_{\rm cl}(n)\rangle/L$ for $n=1$.
(c)~Same as (b) for the running state, showing that the non-propagating 
clusters (full symbols) all become $\nb$-clusters
for large $L$. Averaging $\langle\ldots\rangle$ in (b) and (c) was performed 
over 500 initial particle configurations.}
\label{fig:presoliton_state_f>0}
\end{figure*}

\vspace{1ex}\noindent
The residual free space of an $n$-cluster is
\begin{equation}
r(n)=m(n)-n\sigma\,,
\label{eq:r(n)}
\end{equation}
where
\begin{equation}
m(n)=\lceil n\sigma\rceil
\end{equation}
equals the minimal number of potential wells needed for accommodating
the $n$-cluster.  Here, $\lceil\ldots\rceil$ is the Gaussian bracket
for the ceiling function, i.e.\ $\lceil a\rceil$ is the smallest
integer larger than $a$.

Figure~\ref{fig:presoliton_state_f=0}(b) shows the residual free
spaces $r(n)$ for $n$-clusters composed of particles of size
$\sigma=17/25=0.68$. Applying the free space theorem, we conclude that
only clusters with sizes $n=1$, 4, 7, 10, 13, 16, 19, and 22 can
appear in particle configurations of the presoliton state. They are
marked by red and blue circles in the figure. The clusters in
Fig.~\ref{fig:presoliton_state_f=0}(a) indeed all belong to the
respective sequence of cluster sizes.

Inspite of the different clusters in the configurations of
Fig.~\ref{fig:presoliton_state_f=0}(a), each configuration has the
same number $\Npre=42$ of particles.  In fact, $\Npre$ is unique for a
presoliton state and was derived for infinitesimal force $f=0^+$ and
particle diameters $\sigma=\sigma_{p,q}=p/q$, where integers $p$ and
$q$ are taken to be coprime and $p<q$ ($\sigma<1$). Only for these
$\sigma_{p,q}$, running solitons can occur for $f=0^+$.  It holds
\cite{Antonov/etal:2024}
\begin{equation}
\Npre=\left\lfloor \frac{L}{\mb}\right\rfloor\nb+\left\lfloor \frac{L\ourmod \mb}{\sigma}\right\rfloor\,,
\label{eq:Npre}
\end{equation}
where
\begin{equation}
\nb=q\left\lceil \frac{p^{\varphi(q)-1}}{q} \right\rceil-p^{\varphi(q)-1}
\label{eq:nb}
\end{equation}
is the largest mechanically stable cluster in the presoliton state, and 
\begin{equation}
\mb=m(\nb)=\lceil \nb\sigma\rceil\,.
\label{eq:mb}
\end{equation}
In Eq.~\eqref{eq:Npre}, $\lfloor\ldots\rfloor$ denotes the floor
function, i.e.\ $\lfloor a\rfloor$ is the largest integer smaller than
$a$, and $a \ourmod b=a-\lfloor a/b\rfloor b$ denotes the modulo
operation. The function $\varphi(.)$ in Eq.~\eqref{eq:nb} is Euler's
Phi (totient) function \cite{Abramowitz/Stegun:1965}.

For the example in Fig.~\ref{fig:presoliton_state_f=0}(a), $p=17$ and
$q=25$, and $\nb=22$ from Eq.~\eqref{eq:nb}. This is indeed the
largest stabilizable cluster found in
Fig.~\ref{fig:presoliton_state_f=0}(b). For the system size $L=32$ in
Fig.~\ref{fig:presoliton_state_f=0}(a), we obtain $\Npre=45$ from
Eq.~\eqref{eq:Npre}, in agreement with the particle number in the
different configurations.

The derivation of Eq.~\eqref{eq:Npre} was done by considering a
maximal homogeneous particle configuration of the presoliton state
formed by a sequence of equally spaced largest stabilizable
$\nb$-clusters and by covering the rest part $L-\lfloor L/\mb\rfloor
\mb$ of the system by smaller stabilizable clusters. In fact, the free
space theorem tells us that the $\nb$-cluster has smallest residual
free space $r(\nb$) among the stabilizable clusters and hence is most
efficient when trying to cover the system with largest number of
particles without loosing mechanically stability.  Imagine, to the
contrary, that $\nb>1$ and one would start to fill a system with the
smallest stabilizable clusters of size one, i.e.\ single
particles. Then, after placing a finite number of single particles, it
is possible to combine $\nb$ of them and by this to generate enough
additional free space to place a further particle without loosing
mechanical stability.

Because of this, we expect that when increasing the system size $L$,
only the most efficient covering by $\nb$-clusters will prevail. As a
consequence, the number of $n$-clusters appearing in configurations of
the presoliton state should be extensive for $n=\nb$ only. More
precisely, if we take the set of all mechanically stable particle
configurations of a presoliton state, each of these configurations has
some number $N_{\rm cl}(n)$ of $n$-clusters. Averaging over all
configurations yields the mean number $\langle N_{\rm cl}(n)\rangle$
of $n$-clusters in the presoliton state.  The fraction $\langle N_{\rm
  cl}(\nb)\rangle/L$ then should approach a finite value when
$L\to\infty$, while $\langle N_{\rm cl}(n)\rangle/L\to 0$ for
$n<\nb$. Results in Fig.~\ref{fig:presoliton_state_f=0}(c) agree with
this prediction.

Equations~\eqref{eq:Npre} and \eqref{eq:nb} were derived for
infinitesimal drag force $f=0^+$. For finite $f>0$, one needs to take
into account that for each cluster size $n$, there exists a critical
force $f_c(\sigma,n)$. For $f>f_c(\sigma,n)$, an $n$-cluster looses
mechanical stability. This effect reduces the size of the largest
stabilizable cluster to a value $\nb(f)\le\nb(f=0^+)$, as illustrated
by the blue and red circles in Fig.~\ref{fig:presoliton_state_f=0}(b).
Interestingly, simulation results suggest that the free space theorem
still holds true, i.e.\ all stabilizable clusters can be inferred from
it.

Let us demonstrate this for the same particle diameter $\sigma=0.68$
as considered in Fig.~\ref{fig:presoliton_state_f=0}, choosing a drag
force $f=0.05$, for which
$\nb=10$. Figure~\ref{fig:presoliton_state_f>0} shows results for this
case. In the particle configurations of the presoliton state shown in
Figure~\ref{fig:presoliton_state_f>0}(a), stable clusters of sizes
$1$, 4, 7, and 10 now appear.  As the free space theorem still holds
true, they are the same as identified from the minimal residual free
spaces $r(n)$ of $n$-clusters in
Fig.~\ref{fig:presoliton_state_f=0}(b).  However, we have to exclude
all $n>10$, for which stability is lost due to the larger drag force.
The stabilizable clusters for $f=0.05$ having sizes $n\le\nb(f)=10$
are marked in red in Fig.~\ref{fig:presoliton_state_f=0}(b).

Figure~\ref{fig:presoliton_state_f>0}(b) shows that mean cluster
numbers in the presoliton behave analogously with increasing system
size: $\langle N_{\rm cl}(n)\rangle/L\to0$ for all $n<\nb$ and
$\langle N_{\rm cl}(\nb)\rangle/L\to 1/\mb$.  Remarkably, in the
running state non-propagating clusters have the unique size $\nb$, see
the full symbols in Fig.~\ref{fig:presoliton_state_f>0}(c). Other
clusters (open symbols) are involved in the soliton propagation.

\begin{figure}[t!]
\includegraphics[width=\columnwidth]{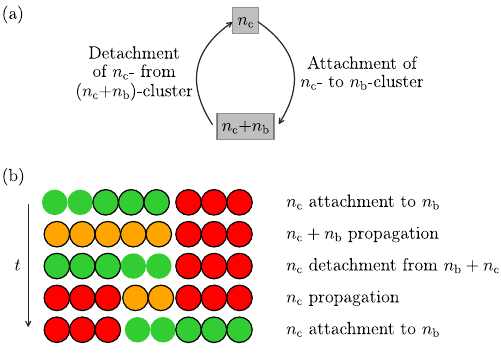}
\caption{Illustration of a soliton propagation. (a) Soliton core
  cluster of size $\nc$ and soliton composite cluster of size
  $\nb+\nc$ form in a periodic process of cluster attachments and
  detachments and give rise to a solitary cluster wave. (b) One period
  of soliton motion, where in the 1st row a core $\nc$-cluster (green without
  circle boundaries) attaches to an $\nb$-cluster (green), the formed
  composite $(\nc\!+\!\nb)$-cluster (orange) thereafter moves in the 2nd row until a
  core $\nc$-cluster detaches from its front in the 3rd row (green without circle
  boundaries). This $\nc$-cluster (orange) moves in the 4th row until attaching 
  to an $\nb$-cluster in the 5th row, which completes the period.}
\label{fig:soliton_propagation}
\end{figure} 

As illustrated in Fig.~\ref{fig:soliton_propagation}(a), soliton
propagates in a periodic process involving attachments and detachments
of a core soliton $\nc$-cluster to and from $\nb$-clusters. In the
sinusoidal potential, one soliton period can be described as follows,
see Fig.~\ref{fig:soliton_propagation}(b).  At the beginning of a
period, a composite $(\nc+\nb)$-cluster (green) is formed by
attachment of an $\nc$-cluster (green circles without black circle
borders) to an $\nb$-cluster (red).  This composite cluster moves
until it becomes unstable and an $\nc$-cluster detaches from its front
part. The detached $\nc$-cluster (orange) thereafter moves until a new
soliton period starts, when the $\nc$-cluster attaches to the next
$\nb$-cluster \footnote{There exists also a variant of the soliton
  propagation, where during the motion of the composite cluster an
  $\nb$-cluster first attaches at its back end and shortly after
  detaches \cite{Antonov/etal:2024}.}.

In Fig.~\ref{fig:soliton_propagation}(c), $\nb=10$ and $\nc=3$, and
accordingly the composite cluster has size $\nb+\nc=13$. Indeed, the
only propagating clusters have size 3 and 13 at large $L$. Only for
small $L$ of order $\mb$, other propagating clusters may appear.

\section{Unit displacement law}
\label{sec:udl}
Particle displacements in states carrying $\Nsol$ solitary cluster
waves are of different types.  Particles forming core $\nc$- and
composite $\nbc$-clusters are translated by a soliton, while those
forming $\nb$-clusters relax towards position of mechanical
equilibria. This relaxation is heterogeneous in space, because it
depends on the distances from solitary waves. Inspite of this
complexity, a remarkably simple unit displacement law holds:\\[1ex]
\hspace*{0.05\columnwidth}\parbox[t]{0.9\columnwidth}{\it After one soliton period, the sum $\Displ$ 
of all particle displacements per soliton is equal to the wavelength of the potential:}
\begin{equation}
\frac{\Displ}{\Nsol}=1\,.
\label{eq:udl}
\end{equation}

\vspace{1ex} We prove this law for a sinusoidal potential in the limit
of infinitesimal drag force in Sec.~\ref{subsec:udl-proof}. For finite
drag forces and an example of a non-sinusoidal potential, we test its
validity by simulations in Sec.~\ref{subsec:udl-simulations}.

\subsection{Proof of unit displacement law for sinusoidal potential and infinitesimal drag force}
\label{subsec:udl-proof}
\begin{figure}[t!]
\includegraphics[width=\columnwidth]{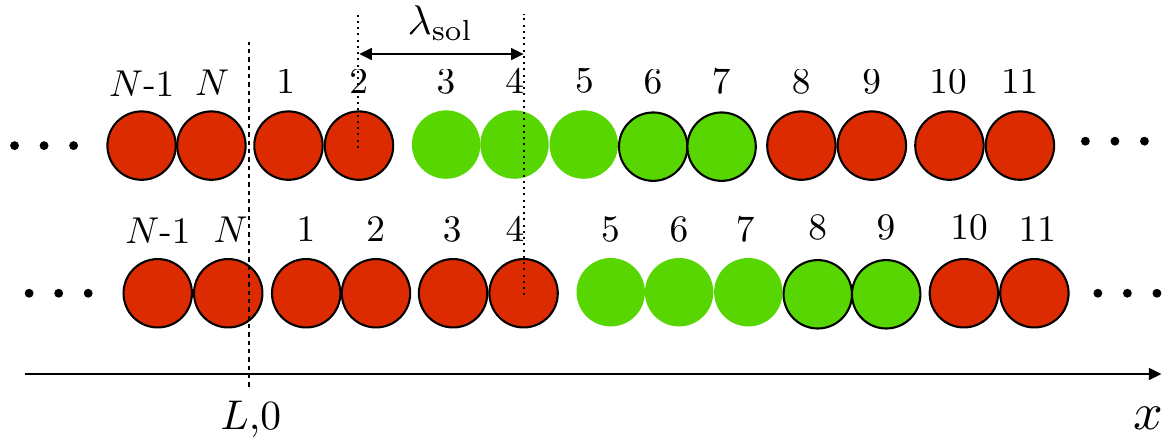}
\caption{Upper row: particle configuration at a time instant, where an
  $\nc$-cluster attaches to an $\nb$-cluster ($\nc=3$, $\nb=2$).
  Basic $\nb$-clusters are marked in red.  The $\nbc$-cluster marked
  in green is forming in the attachment process.  Lower row:
  Equivalent configuration after one soliton period, where particles
  with index $i$ correspond to particles with index $i-\nb$ (modulo
  $N$) in the upper row. The distance between corresponding particles
  is equal to the soliton period $\lambdasol$, see
  Eq.~\eqref{eq:xi-xiprime}.  Particles of the attaching clusters in
  the two equivalent configurations are depicted without black circle
  borders.  The vertical dashed line indicates the origin $x=0$ of the
  $x$-axis, which is identical to $x=L$ due to periodic boundary
  conditions.}
\label{fig:illustration_for_udl_proof}
\end{figure}

For a steady state carrying $\Nsol$ solitons, the sum of particle
displacements in one soliton period can be calculated by considering
two subsequent time instants, where an $\nc$-cluster attaches to an
$\nb$-cluster. Two particle configurations at such instants are
illustrated in Fig.~\ref{fig:illustration_for_udl_proof}. In the
initial particle configuration in the upper row, we labeled the
particles from 1 to $N$, starting with the particle right from the
origin $x=0$. The particle positions in this configurations are $x_i$,
$i=1,\ldots, N$, and in the final configuration after one soliton
period (lower row) they are $x_i'$. The sum of particle displacements
is
\begin{equation}
\Displ=\sum_{i=1}^N (x_i'-x_i)\,.
\label{eq:Delta-def}
\end{equation}
The initial and final configurations are equivalent. For example,
particles 1, 2, 3, 4 in the final configuration in
Fig.~\ref{fig:illustration_for_udl_proof} correspond to particles
$N\!-\!1$, $N$, 1, 2 in the initial configuration. Hence, if we shift
the particle indices of the final configuration by $-\nb=-2$, we
obtain the corresponding particle indices (modulo $N$) in the initial
configuration.  As a consequence of the equivalence, the particle
positions are related as
\begin{equation}
x_i'=\left\{\begin{array}{cc}
x_{i-\nb+N}+\lambdasol-L\,, & i=1,\ldots,\nb\,,\\[1ex]
x_{i-\nb}+\lambdasol\,, & i=\nb+1,\ldots,N\,.
\end{array}\right.
\label{eq:xi-xiprime}
\end{equation}
The periodic boundary conditions are taken into account by taking
particle indices modulo $N$ and particle positions modulo $L$.

Inserting Eq.~\eqref{eq:xi-xiprime} into Eq.~\eqref{eq:Delta-def}, we
obtain
\begin{equation}
\Displ=N\lambdasol-\nb L\,,
\label{eq:D1}
\end{equation}
i.e.\ $\Displ$ must be an integer. 

We now show that $\Displ$ is an integer multiple of $\Nsol$ by taking
into account two conditions: First, the total number of particles in
$\Nb$ basic clusters composed of $\nb$ particles and $\Nsol$ soliton
clusters composed of $\nsol=n_{\rm c}+n_{\rm b}$ particles must equal
the particle number $N$,
\begin{equation}
\Nb\nb+\Nsol\nsol=N\,.
\label{eq:Nrule}
\end{equation}
Second, in the running state, all potential wells must be occupied by
clusters to enable soliton propagation by sequential attachments and
detachments processes, see Fig.~\ref{fig:soliton_propagation}.
Accordingly, the sum of potential wells occupied by the clusters gives
the system size $L$,
\begin{equation}
\Nb\mb+\Nsol\msol=L\,.
\label{eq:Lrule}
\end{equation}
Here, $\msol$ is the number of potential wells needed for containing
one soliton and $\mb$ is given in Eq.~\eqref{eq:mb}.

For the sinusoidal potential, the distance $\lambda_{\rm sol}$
traveled by a soliton in one period is equal to $\mb$
\cite{Antonov/etal:2024}. This can be intuitively understood from the
soliton propagation discussed in
Sec.~\ref{sec:presoliton-running-state}: after completing one soliton
period, an $\nb$-cluster is created that is accommodated by $\mb$
wells -- see the example in Fig.~\ref{fig:illustration_for_udl_proof},
where the 2-cluster composed of particles 3 and 4 in the lower row is
created by the soliton movement. Associated with this process is a
displacement of the two equivalent particle configurations by
$\lambdasol$.

Taking $\lambda_{\rm sol}=\mb$, and inserting $N$ and $L$ from
Eqs.~\eqref{eq:Nrule} and \eqref{eq:Lrule} into Eq.~\eqref{eq:D1}, we
obtain
\begin{equation}
\Displ=\Nsol(\nsol\mb-\nb\msol)\,.
\label{eq:D2}
\end{equation}

In the limit of infinitesimal drag force, where soliton propagation
can occur only for certain rational particle sizes $\sigma_{p,q}=p/q$,
$\nsol=q$, $\mb=\lceil \nb\sigma_{p,q}\rceil=\nb(p/q)+1/q$, and
$\msol=\lceil\nsol\sigma_{p,q}\rceil=p$ \cite{Antonov/etal:2024}. This
implies
\begin{equation}
\nsol\mb-\nb\msol=q\left(\frac{p}{q}\,\nb+\frac{1}{q}\right)-\nb p=1\,.
\label{eq:msol-zero-force}
\end{equation}
Hence, Eq.~\eqref{eq:udl} holds for infinitesimal drag force.

\subsection{Tests of unit displacement law by simulations}
\label{subsec:udl-simulations}
\begin{figure}[t!]
\centering
\includegraphics[width=\columnwidth]{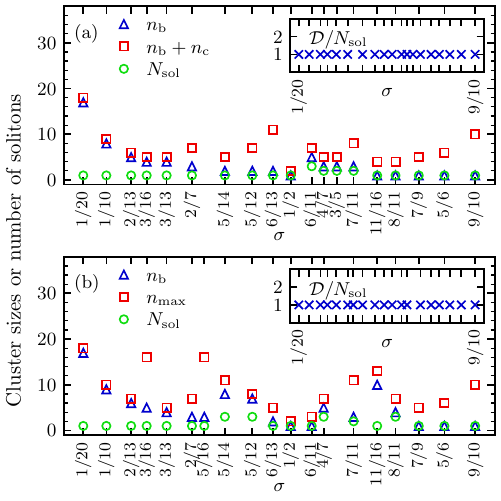}
\caption{Verification of UDL by simulations. Sizes of
  largest ($\nb+\nc$ or $n_{\rm max}$) and basic stable clusters
  ($\nb$) for different particle diameters $\sigma=p/q$,
  together with number $\Nsol$ of solitons. Part (a)
  shows the results for a sinusoidal potential and part (b) for the
  triangle wave potential \eqref{eq:triangle-wave-potential} with
  smoothing parameter $\delta=0.02$. In both cases the system size is
  $L=30$ and the drag force $f=0.2$. Inspite of strong and irregular
  variation of the cluster sizes and number of solitons, the total sum $ \mathcal{D}$ of particle
  displacements in one soliton period is equal to $\Nsol$, see
  insets.}
\label{fig:udl-simulation}
\end{figure}

For finite $f=0.2$ in the sinusoidal potential, simulation results for
$\nb$, $\nb+\nc$ and $\Nsol$ are shown for various $\sigma_{p,q}$ in
Fig.~\ref{fig:udl-simulation}(a). The system size is $L=30$ and
particle numbers are $N=\Npre+1$, i.e.\ we added one particle to the
presoliton state. While $\nb$ and $\nb+\nc$, and the number of
solitons vary strongly with $\sigma$ and without obvious regularity,
the sum of particle displacements per soliton $\mathcal{D}/\Nsol$ is
always one, as shown in the inset of Fig.~\ref{fig:udl-simulation}(a).

For the same system size $L=30$ and drag force $f=0.2$, we display in
Fig.~\ref{fig:udl-simulation}(b) corresponding simulation results for
the triangle wave potential \eqref{eq:triangle-wave-potential} with
smoothing parameter $ \delta=0.02$. The soliton propagation can be
more complex in this potential, involving more cluster types. Instead
of $ \nb+\nc$, we show the size $n_{\rm max}$ of the largest cluster
appearing in the soliton propagation. Analogous to the sinusoidal
potential, $\nb$, $n_{\rm max}$ and $\Nsol$ vary strongly and in an
irregular manner with $\sigma$, whereas $\mathcal{D}/\Nsol=1$ in all
cases.

We would like to point out that the individual particle displacements
in one soliton period are different. If a single particle is part of a
cluster relaxing towards a position of mechanical equilibrium, its
displacement depends on how much the cluster is away from its point of
stability. If a single particle is involved in the soliton
propagation, its displacement depends on the soliton cluster of which
it is part of and how much this cluster becomes displaced in one
soliton period. Hence, there is a strong heterogeneity of individual
particle displacements but irrespective of this feature, the sum of
all particle displacements obeys the simple rule \eqref{eq:udl}.

\section{Implications of UDL}
\label{sec:consequences}
The UDL allows us to determine the spatial extension $\msol$ of a
soliton, and to derive particle currents mediated by solitary cluster
waves. Furthermore, it determines $\nc$ up to an integer multiple of
$\nb$. When adding a single particle to the system, it relates the
increase $\delta\Nsol$ of soliton number and decrease $\-\delta\Nb$ of
number of $\nb$-clusters to $\mb$ and $\msol$.

\subsection{Spatial extension of solitons }
\label{subsec:nc}
For the soliton extension, 
we obtain from Eqs.~ \eqref{eq:udl} and \eqref{eq:D2}
\begin{equation}
\msol=\frac{1}{\nb}(\nsol\,\mb-1)\,.
\label{eq:msol}
\end{equation}
In experiments, this expression can be evaluated by determining the
size $\nb$ of the relaxing basic cluster and the maximal number
$\nsol$ of particles involved in the soliton propagation.

It is important to note that $\msol$ is not equal to $\lceil
\nsol\sigma\rceil=\lceil (\nb+\nc)\sigma \rceil$ in general.  For
example, in case of the sinusoidal potential and particle diameter
$\sigma=0.61$, $\nb=3$ and $\nc=2$ for $f=0.3$.  This yields the
soliton extension $\msol=3$ according to Eq.~\eqref{eq:msol}, while
the number of wells accommodating an $\nsol$-cluster is $\lceil
\nsol\sigma\rceil=\lceil (\nb+\nc)\sigma\rceil=4$.

\subsection{Soliton-mediated particle currents}
\label{subsec:currents}
From the UDL follows a simple expression for the particle current $J$:
the mean particle velocity is $\mathcal{D}/N\tausol$, where $\tausol$
is the time period of the solitary wave propagation, i.e.\ the time
needed for the soliton to move a distance
$\lambdasol=\mb$. Multiplying this mean velocity with the number
density $N/L$ yields
\begin{equation}
J=\frac{N}{L}\frac{\mathcal{D}}{N\tausol}=\frac{\Nsol}{L\tausol}\,.
\label{eq:J}
\end{equation}
For the sinusoidal potential, explicit expressions for $\Nsol$ and
$\tausol$ in terms of $\sigma$, $L$, $\nb$, and $\nc$ are given in
\cite{Antonov/etal:2024}.

Also, another derivation of the particle current was provided,
yielding the seemingly unrelated result
\begin{equation}
J=\frac{\Nsol d}{\nb L+\Nsol d}\frac{N}{L}\frac{\mb}{\tausol}\,,
\label{eq:Jformer}
\end{equation}
where $d$ is the displacement of each particle after a single soliton
has traveled a distance $\nb L+d$ \footnote{In
  Ref.~\cite{Antonov/etal:2024}, there is $\nb/\gcd(\nb,\nc)$ in the
  denominator of Eq.~\eqref{eq:Jformer} instead of $\nb$. However, it
  was shown that $\nb$ and $\nb+\nc$ are coprime and hence $\nb$ and
  $\nc$ are coprime also, i.e.\ their greatest common divisor
  satisfies $\gcd(\nb,\nc)=\gcd(\nb+\nc,\nc)=1$.}.

Comparing Eqs.~\eqref{eq:J} and \eqref{eq:Jformer} yields
\begin{equation}
\nb+\frac{\Nsol}{L}d=\frac{N}{L}\mb d\,.
\label{eq:ddet}
\end{equation}
The equation must hold true for all system sizes $L$. We thus can
increase $L$ by adding $M\mb$ potential wells, where in each of the
$M$ intervals $\mb$ we place one basic stable $\nb$-cluster. This way,
we extend the system by only adding stable clusters, i.e.\ $\Nsol$
does not change. Accordingly, be letting $L\to L+M\mb$ and $N\to
N+M\nb$ with $M\to\infty$, we obtain $N/L\to \nb/\mb$, $\Nsol/L\to0$,
and accordingly $\nb=\nb d$. Hence,
\begin{equation}
d=1\,.
\label{eq:d=1}
\end{equation}
This equation can be considered as an alternative statement of the UDL. 

\subsection{Changes of numbers of solitons and\\ basic stable clusters}
Adding one particle to a running state, increases the number of
solitons by $\delta\Nsol$ and decreases the number of stable clusters
by $\delta\Nb$. It turns out that
\begin{align}
\delta\Nsol&=\mb\,,
\label{eq:Nsol-mb}\\
\delta\Nb&=\msol\,.
\label{eq:Nb-msol}
\end{align}
Equation~\eqref{eq:Nsol-mb} follows by setting $d=1$ in
Eq.~\eqref{eq:ddet} and calculating the change of $\Nsol$ when
incrementing $N$ by one. The result for $\delta\Nb$ then follows from
the space-filling condition \eqref{eq:Lrule} that implies
$-\delta\Nb\mb+\delta\Nsol\msol=0$, which, when using
$\delta\Nsol=\mb$, gives Eq.~\eqref{eq:Nb-msol}.

Based on Eq.~\eqref{eq:Nsol-mb} it is possible to design running
states with controlled stepwise increase of solitons upon adding
particles. A specific $\mb=\delta\Nsol$ value can be realized by
choosing $\sigma$ and $f$ to yield a corresponding size $\nb$ of the
basic stable cluster.

Equation~\eqref{eq:Nb-msol} is useful to determine soliton
sizes in experiments, where the detailed propagation process by
cluster attachments and detachments can be difficult to resolve in
view of the dense arrangement of the soliton clusters. By counting the
number of clearly separated $\nb$-clusters in running states, one can
obtain $\msol=\delta\Nb$ in a feasible manner.

\subsection{Soliton core cluster}
While for the sinusoidal potential at finite $f>0$ it is possible to
determine $\nb$ quickly based on the cluster of maximal size with
translational stability and by checking fragmentation stability for
this and all clusters of smaller size, an efficient method for
determining $\nc$ is not yet available.  For determining $\nc$, we
need to evaluate for increasing trial sizes $\nc'$ whether in two
parts of a well defined interval of width $r(\nb)$, a core and
composite soliton can propagate. This demanding procedure was
discussed in Sec.~6 of Ref.~\cite{Antonov/etal:2024}.  We here show
that the trial sizes of the core soliton cluster can be restricted to
numbers incremented by $\nb$. 

When increasing $N$ by one, Eq.~\eqref{eq:Nrule} implies
$-\nb\delta\Nb+\nsol\delta\Nsol=1$, which after inserting
$\nsol=\nb+\nc$ and $\delta\Nsol=\mb$ gives a linear Diophantine
equation in the two variables $\nc$ and $\delta\Nb$,
\begin{equation}
\nb\delta\Nb-\mb\nc=\nb\mb-1\,.
\end{equation}
The solutions for $\nc$ are \cite{Vorobyov:1980}
\begin{align}
\nc=&\left\lceil\frac{1+(\nb\mb\!-\!1)\mb^{\varphi(\nb')-1}}{\nb}\right\rceil\nb
\label{eq:nc-j}\\
&{}-(\nb\mb\!-\!1)\,\mb^{\varphi(\nb')-1}+j\nb\,,\hspace{1em}j=0,1\ldots
\nonumber
\end{align}
Here, $\nb'=\nb/\gcd(\nb,\mb)$, where $\gcd(\nb,\mb)$ is the greatest
common divisor of $\nb$ and $\mb$.

The $\nc$ obtained from simulation results of
Fig.~\ref{fig:udl-simulation}(a) agree with Eq.~\eqref{eq:nc-j} for different $j$.


For infinitesimal force, where running states are possible for
rational $\sigma=p/q$ only, the core $\nc$-cluster has size
$\nc=q-\nb$ and a $q$-cluster can move barrier-free as in a flat
potential landscape, as long as it does not fragment.  The smaller
clusters of size $n$, $n=\ldots,q-1$, cannot move barrier-free and all
have a different residual free space
$r(n)\in\{1/q,2/q,\ldots,(q-1)/q\}$.  The $\nc$-cluster with
$\nc=q-\nb$ has largest residual free space, i.e.\ $r(\nc)=(q-1)/q$.
This follows from the fact that $\nb$ has smallest residual free space
$r(\nb)=1/q$:
\begin{align}
r(q\!-\!\nb)&=\left\lceil (q\!-\!\nb)\frac{p}{q}\right\rceil-(q\!-\!\nb)\frac{p}{q}=\left\lceil -\nb\frac{p}{q}\right\rceil+\nb\frac{p}{q}\nonumber\\
&=-\left\lfloor \nb\frac{p}{q}\right\rfloor+\nb\frac{p}{q}=1-\left\lceil \nb\frac{p}{q}\right\rceil+\nb\frac{p}{q}\nonumber\\
&=1-r(\nb)=1-\frac{1}{q}=\frac{q-1}{q}\,.
\end{align}

Motivated by this result, we studied the residual free space of
$\nc$-clusters at finite $f$ also, and found in all simulation results
that $\nc$ fulfils the following maximal residual free space
property:\\[0.5ex]
\hspace*{0.05\columnwidth}\parbox[t]{0.9\columnwidth}{\it  
The residual free space $r(n)$ of all clusters $n<\nc$ is smaller than $r(\nc)$.}

\vspace{0.5ex} Combining this property with the selection procedure
\eqref{eq:nc-j}, frequently gives a unique solution for $\nc$
already. In cases, where it does not, the computational effort for
determining $\nc$ is reduced strongly.

One can improve the $\nc$-determination based on purely geometric
considerations also by calculating that $j$ in the selection procedure
\eqref{eq:nc-j}, for which $r(\nc)$ is maximal; we also required
$\nc<q-\nb$ for $\sigma=p/q$. This method gives agreement with the
simulated $\nc$ in Fig.~\ref{fig:udl-simulation}(a) except for
$\sigma=3/13$, 6/13, 7/11, and 8/11.

A full agreement with the simulation results is obtained by calculating that $j$ in the selection procedure
\eqref{eq:nc-j}, for which the $\nc$- and the $(\nb+\nc)$-cluster at the point of attachment 
of the $\nc$- to the $\nb$-cluster (cf.\ Fig.~\ref{fig:soliton_propagation}) do not fragment.

\section{Conclusions}
\label{sec:conclusions}
We have reported new important properties of the recently discovered
solitary cluster waves.  First, we have demonstrated that particle
configurations of presoliton states can be formed by stable clusters
with different sizes for small system lengths $L$. However, only the
mean number of stable clusters with largest size $\nb$ grows
proportionally with $L$. Differently speaking, in presoliton states
only the mean cluster number $\langle N(\nb)\rangle$ is extensive in
system size. In soliton-carrying running states, there is uniqueness
of non-propagating clusters: all of them are $\nb$-clusters.

Our next key result is the UDL, which states that the sum of all
particle displacements per soliton in one soliton period is equal to
one wavelength of the periodic potential.  We derived this law for the
sinusoidal potential and for infinitesimal drag force $f=0^+$, and
confirmed it by simulations for finite $f>0$ for both a sinusoidal and
a non-sinusoidal one.

We discussed various consequences of this law. It gives a simple
expression for the soliton extension, simplifies strongly exact
results for soliton-mediated currents as well as calculations of the
soliton core cluster size.  It implies that the increase in soliton
number and decrease in number of $\nb$-clusters upon adding a particle
to a running state is given by the number of accommodating wells of
the $\nb$-cluster and the soliton extension.  The respective relations
can be useful to extract soliton properties from experimental
observations like those in Refs.~\cite{Bohlein/etal:2012,
  Bohlein/Bechinger:2012, Tierno/Fischer:2014, Tierno/etal:2014,
  Juniper/etal:2015, Cao/etal:2019b, Stoop/etal:2020,
  Mirzaee-Kakhki/etal:2020, Leyva/etal:2022,
  Cereceda-Lopez/etal:2023}.

In applications of the UDL to soliton-carrying states under
time-periodic driving, it is possible that equivalent particle
configurations occur after a minimal number $p>1$ of periods of the
driving. The law then generalizes to that the sum of particle
displacements after $p$ periods of soliton motion equals $p$ potential
wavelengths.

So far we have considered periodic potentials with point symmetry in
one dimension. An open question is whether the properties of
presoliton and soliton-carrying states remain valid in asymmetric
periodic potentials allowing for ratcheting, and in higher
dimensions. Even if cluster attachments and detachments involved in
the soliton propagation can be more versatile then, the UDL may 
still hold true. This would allow one to quantify
currents in cluster-mediated particle transport despite higher
complexity of the soliton propagation process.

\begin{acknowledgments}
We thank P.~Tierno for discussions on soliton dynamics in experiments,
and the Czech Science Foundation (Project No.\ 23-09074L) and the
Deutsche Forschungsgemeinschaft (Project No.\ 521001072) for financial
support.  The use of a high-performance computing cluster funded by
the Deutsche Forschungsgemeinschaft is gratefully acknowledged
(Project No.\ 456666331).
\end{acknowledgments}


\begin{thebibliography}{32}%
\makeatletter
\providecommand \@ifxundefined [1]{%
 \@ifx{#1\undefined}
}%
\providecommand \@ifnum [1]{%
 \ifnum #1\expandafter \@firstoftwo
 \else \expandafter \@secondoftwo
 \fi
}%
\providecommand \@ifx [1]{%
 \ifx #1\expandafter \@firstoftwo
 \else \expandafter \@secondoftwo
 \fi
}%
\providecommand \natexlab [1]{#1}%
\providecommand \enquote  [1]{``#1''}%
\providecommand \bibnamefont  [1]{#1}%
\providecommand \bibfnamefont [1]{#1}%
\providecommand \citenamefont [1]{#1}%
\providecommand \href@noop [0]{\@secondoftwo}%
\providecommand \href [0]{\begingroup \@sanitize@url \@href}%
\providecommand \@href[1]{\@@startlink{#1}\@@href}%
\providecommand \@@href[1]{\endgroup#1\@@endlink}%
\providecommand \@sanitize@url [0]{\catcode `\\12\catcode `\$12\catcode
  `\&12\catcode `\#12\catcode `\^12\catcode `\_12\catcode `\%12\relax}%
\providecommand \@@startlink[1]{}%
\providecommand \@@endlink[0]{}%
\providecommand \url  [0]{\begingroup\@sanitize@url \@url }%
\providecommand \@url [1]{\endgroup\@href {#1}{\urlprefix }}%
\providecommand \urlprefix  [0]{URL }%
\providecommand \Eprint [0]{\href }%
\providecommand \doibase [0]{https://doi.org/}%
\providecommand \selectlanguage [0]{\@gobble}%
\providecommand \bibinfo  [0]{\@secondoftwo}%
\providecommand \bibfield  [0]{\@secondoftwo}%
\providecommand \translation [1]{[#1]}%
\providecommand \BibitemOpen [0]{}%
\providecommand \bibitemStop [0]{}%
\providecommand \bibitemNoStop [0]{.\EOS\space}%
\providecommand \EOS [0]{\spacefactor3000\relax}%
\providecommand \BibitemShut  [1]{\csname bibitem#1\endcsname}%
\let\auto@bib@innerbib\@empty
\bibitem [{\citenamefont {Karnik}\ \emph {et~al.}(2007)\citenamefont {Karnik},
  \citenamefont {Duan}, \citenamefont {Castelino}, \citenamefont {Daiguji},\
  and\ \citenamefont {Majumdar}}]{Karnik/etal:2007}%
  \BibitemOpen
  \bibfield  {author} {\bibinfo {author} {\bibfnamefont {R.}~\bibnamefont
  {Karnik}}, \bibinfo {author} {\bibfnamefont {C.}~\bibnamefont {Duan}},
  \bibinfo {author} {\bibfnamefont {K.}~\bibnamefont {Castelino}}, \bibinfo
  {author} {\bibfnamefont {H.}~\bibnamefont {Daiguji}},\ and\ \bibinfo {author}
  {\bibfnamefont {A.}~\bibnamefont {Majumdar}},\ }\bibfield  {title} {\bibinfo
  {title} {Rectification of ionic current in a nanofluidic diode},\ }\href
  {https://doi.org/10.1021/nl062806o} {\bibfield  {journal} {\bibinfo
  {journal} {Nano Lett.}\ }\textbf {\bibinfo {volume} {7}},\ \bibinfo {pages}
  {547} (\bibinfo {year} {2007})}\BibitemShut {NoStop}%
\bibitem [{\citenamefont {Misiunas}\ and\ \citenamefont
  {Keyser}(2019)}]{Misiunas/Keyser:2019}%
  \BibitemOpen
  \bibfield  {author} {\bibinfo {author} {\bibfnamefont {K.}~\bibnamefont
  {Misiunas}}\ and\ \bibinfo {author} {\bibfnamefont {U.~F.}\ \bibnamefont
  {Keyser}},\ }\bibfield  {title} {\bibinfo {title} {Density-dependent speed-up
  of particle transport in channels},\ }\href
  {https://doi.org/10.1103/PhysRevLett.122.214501} {\bibfield  {journal}
  {\bibinfo  {journal} {Phys. Rev. Lett.}\ }\textbf {\bibinfo {volume} {122}},\
  \bibinfo {pages} {214501} (\bibinfo {year} {2019})}\BibitemShut {NoStop}%
\bibitem [{\citenamefont {Su}\ \emph {et~al.}(2022)\citenamefont {Su},
  \citenamefont {Zhang}, \citenamefont {Peng}, \citenamefont {Guo},
  \citenamefont {Liu}, \citenamefont {Fu}, \citenamefont {Yao}, \citenamefont
  {Du}, \citenamefont {Du},\ and\ \citenamefont {Xue}}]{Su/etal:2022}%
  \BibitemOpen
  \bibfield  {author} {\bibinfo {author} {\bibfnamefont {S.}~\bibnamefont
  {Su}}, \bibinfo {author} {\bibfnamefont {Y.}~\bibnamefont {Zhang}}, \bibinfo
  {author} {\bibfnamefont {S.}~\bibnamefont {Peng}}, \bibinfo {author}
  {\bibfnamefont {L.}~\bibnamefont {Guo}}, \bibinfo {author} {\bibfnamefont
  {Y.}~\bibnamefont {Liu}}, \bibinfo {author} {\bibfnamefont {E.}~\bibnamefont
  {Fu}}, \bibinfo {author} {\bibfnamefont {H.}~\bibnamefont {Yao}}, \bibinfo
  {author} {\bibfnamefont {J.}~\bibnamefont {Du}}, \bibinfo {author}
  {\bibfnamefont {G.}~\bibnamefont {Du}},\ and\ \bibinfo {author}
  {\bibfnamefont {J.}~\bibnamefont {Xue}},\ }\bibfield  {title} {\bibinfo
  {title} {Multifunctional graphene heterogeneous nanochannel with
  voltage-tunable ion selectivity},\ }\href
  {https://doi.org/10.1038/s41467-022-32590-9} {\bibfield  {journal} {\bibinfo
  {journal} {Nat. Commun.}\ }\textbf {\bibinfo {volume} {13}},\ \bibinfo
  {pages} {4894} (\bibinfo {year} {2022})}\BibitemShut {NoStop}%
\bibitem [{\citenamefont {Bressloff}\ and\ \citenamefont
  {Newby}(2013)}]{Bressloff/Newby:2013}%
  \BibitemOpen
  \bibfield  {author} {\bibinfo {author} {\bibfnamefont {P.~C.}\ \bibnamefont
  {Bressloff}}\ and\ \bibinfo {author} {\bibfnamefont {J.~M.}\ \bibnamefont
  {Newby}},\ }\bibfield  {title} {\bibinfo {title} {Stochastic models of
  intracellular transport},\ }\href {https://doi.org/10.1103/RevModPhys.85.135}
  {\bibfield  {journal} {\bibinfo  {journal} {Rev. Mod. Phys.}\ }\textbf
  {\bibinfo {volume} {85}},\ \bibinfo {pages} {135} (\bibinfo {year}
  {2013})}\BibitemShut {NoStop}%
\bibitem [{\citenamefont {Xiao}\ \emph {et~al.}(2003)\citenamefont {Xiao},
  \citenamefont {Greaney},\ and\ \citenamefont {Chrzan}}]{Xiao/etal:2003}%
  \BibitemOpen
  \bibfield  {author} {\bibinfo {author} {\bibfnamefont {W.}~\bibnamefont
  {Xiao}}, \bibinfo {author} {\bibfnamefont {P.~A.}\ \bibnamefont {Greaney}},\
  and\ \bibinfo {author} {\bibfnamefont {D.~C.}\ \bibnamefont {Chrzan}},\
  }\bibfield  {title} {\bibinfo {title} {Adatom transport on strained
  {C}u(001): Surface crowdions},\ }\href
  {https://doi.org/10.1103/PhysRevLett.90.156102} {\bibfield  {journal}
  {\bibinfo  {journal} {Phys. Rev. Lett.}\ }\textbf {\bibinfo {volume} {90}},\
  \bibinfo {pages} {156102} (\bibinfo {year} {2003})}\BibitemShut {NoStop}%
\bibitem [{\citenamefont {Korda}\ \emph {et~al.}(2002)\citenamefont {Korda},
  \citenamefont {Taylor},\ and\ \citenamefont {Grier}}]{Korda/etal:2002}%
  \BibitemOpen
  \bibfield  {author} {\bibinfo {author} {\bibfnamefont {P.~T.}\ \bibnamefont
  {Korda}}, \bibinfo {author} {\bibfnamefont {M.~B.}\ \bibnamefont {Taylor}},\
  and\ \bibinfo {author} {\bibfnamefont {D.~G.}\ \bibnamefont {Grier}},\
  }\bibfield  {title} {\bibinfo {title} {Kinetically locked-in colloidal
  transport in an array of optical tweezers},\ }\href
  {https://doi.org/10.1103/PhysRevLett.89.128301} {\bibfield  {journal}
  {\bibinfo  {journal} {Phys. Rev. Lett.}\ }\textbf {\bibinfo {volume} {89}},\
  \bibinfo {pages} {128301} (\bibinfo {year} {2002})}\BibitemShut {NoStop}%
\bibitem [{\citenamefont {Bohlein}\ \emph {et~al.}(2012)\citenamefont
  {Bohlein}, \citenamefont {Mikhael},\ and\ \citenamefont
  {Bechinger}}]{Bohlein/etal:2012}%
  \BibitemOpen
  \bibfield  {author} {\bibinfo {author} {\bibfnamefont {T.}~\bibnamefont
  {Bohlein}}, \bibinfo {author} {\bibfnamefont {J.}~\bibnamefont {Mikhael}},\
  and\ \bibinfo {author} {\bibfnamefont {C.}~\bibnamefont {Bechinger}},\
  }\bibfield  {title} {\bibinfo {title} {Observation of kinks and antikinks in
  colloidal monolayers driven across ordered surfaces},\ }\href
  {https://doi.org/10.1038/nmat3204} {\bibfield  {journal} {\bibinfo  {journal}
  {Nat. Mater.}\ }\textbf {\bibinfo {volume} {11}},\ \bibinfo {pages} {126}
  (\bibinfo {year} {2012})}\BibitemShut {NoStop}%
\bibitem [{\citenamefont {Bohlein}\ and\ \citenamefont
  {Bechinger}(2012)}]{Bohlein/Bechinger:2012}%
  \BibitemOpen
  \bibfield  {author} {\bibinfo {author} {\bibfnamefont {T.}~\bibnamefont
  {Bohlein}}\ and\ \bibinfo {author} {\bibfnamefont {C.}~\bibnamefont
  {Bechinger}},\ }\bibfield  {title} {\bibinfo {title} {Experimental
  observation of directional locking and dynamical ordering of colloidal
  monolayers driven across quasiperiodic substrates},\ }\href
  {https://doi.org/10.1103/PhysRevLett.109.058301} {\bibfield  {journal}
  {\bibinfo  {journal} {Phys. Rev. Lett.}\ }\textbf {\bibinfo {volume} {109}},\
  \bibinfo {pages} {058301} (\bibinfo {year} {2012})}\BibitemShut {NoStop}%
\bibitem [{\citenamefont {Tierno}\ and\ \citenamefont
  {Fischer}(2014)}]{Tierno/Fischer:2014}%
  \BibitemOpen
  \bibfield  {author} {\bibinfo {author} {\bibfnamefont {P.}~\bibnamefont
  {Tierno}}\ and\ \bibinfo {author} {\bibfnamefont {T.~M.}\ \bibnamefont
  {Fischer}},\ }\bibfield  {title} {\bibinfo {title} {Excluded volume causes
  integer and fractional plateaus in colloidal ratchet currents},\ }\href
  {https://doi.org/10.1103/PhysRevLett.112.048302} {\bibfield  {journal}
  {\bibinfo  {journal} {Phys. Rev. Lett.}\ }\textbf {\bibinfo {volume} {112}},\
  \bibinfo {pages} {048302} (\bibinfo {year} {2014})}\BibitemShut {NoStop}%
\bibitem [{\citenamefont {Tierno}\ \emph {et~al.}(2014)\citenamefont {Tierno},
  \citenamefont {Johansen},\ and\ \citenamefont {Fischer}}]{Tierno/etal:2014}%
  \BibitemOpen
  \bibfield  {author} {\bibinfo {author} {\bibfnamefont {P.}~\bibnamefont
  {Tierno}}, \bibinfo {author} {\bibfnamefont {T.~H.}\ \bibnamefont
  {Johansen}},\ and\ \bibinfo {author} {\bibfnamefont {T.~M.}\ \bibnamefont
  {Fischer}},\ }\bibfield  {title} {\bibinfo {title} {Fast and rewritable
  colloidal assembly via field synchronized particle swapping},\ }\href
  {https://doi.org/10.1063/1.4874839} {\bibfield  {journal} {\bibinfo
  {journal} {Appl. Phys. Lett.}\ }\textbf {\bibinfo {volume} {104}},\ \bibinfo
  {pages} {174102} (\bibinfo {year} {2014})}\BibitemShut {NoStop}%
\bibitem [{\citenamefont {Juniper}\ \emph {et~al.}(2015)\citenamefont
  {Juniper}, \citenamefont {Straube}, \citenamefont {Besseling}, \citenamefont
  {Aarts},\ and\ \citenamefont {Dullens}}]{Juniper/etal:2015}%
  \BibitemOpen
  \bibfield  {author} {\bibinfo {author} {\bibfnamefont {M.~P.~N.}\
  \bibnamefont {Juniper}}, \bibinfo {author} {\bibfnamefont {A.~V.}\
  \bibnamefont {Straube}}, \bibinfo {author} {\bibfnamefont {R.}~\bibnamefont
  {Besseling}}, \bibinfo {author} {\bibfnamefont {D.~G. A.~L.}\ \bibnamefont
  {Aarts}},\ and\ \bibinfo {author} {\bibfnamefont {R.~P.~A.}\ \bibnamefont
  {Dullens}},\ }\bibfield  {title} {\bibinfo {title} {Microscopic dynamics of
  synchronization in driven colloids},\ }\href
  {https://doi.org/10.1038/ncomms8187} {\bibfield  {journal} {\bibinfo
  {journal} {Nat. Commun.}\ }\textbf {\bibinfo {volume} {6}},\ \bibinfo {pages}
  {7187} (\bibinfo {year} {2015})}\BibitemShut {NoStop}%
\bibitem [{\citenamefont {Cao}\ \emph {et~al.}(2019)\citenamefont {Cao},
  \citenamefont {Panizon}, \citenamefont {Vanossi}, \citenamefont {Manini},\
  and\ \citenamefont {Bechinger}}]{Cao/etal:2019b}%
  \BibitemOpen
  \bibfield  {author} {\bibinfo {author} {\bibfnamefont {X.}~\bibnamefont
  {Cao}}, \bibinfo {author} {\bibfnamefont {E.}~\bibnamefont {Panizon}},
  \bibinfo {author} {\bibfnamefont {A.}~\bibnamefont {Vanossi}}, \bibinfo
  {author} {\bibfnamefont {N.}~\bibnamefont {Manini}},\ and\ \bibinfo {author}
  {\bibfnamefont {C.}~\bibnamefont {Bechinger}},\ }\bibfield  {title} {\bibinfo
  {title} {Orientational and directional locking of colloidal clusters driven
  across periodic surfaces},\ }\href
  {https://doi.org/10.1038/s41567-019-0515-7} {\bibfield  {journal} {\bibinfo
  {journal} {Nat. Phys.}\ }\textbf {\bibinfo {volume} {15}},\ \bibinfo {pages}
  {776} (\bibinfo {year} {2019})}\BibitemShut {NoStop}%
\bibitem [{\citenamefont {Stoop}\ \emph {et~al.}(2020)\citenamefont {Stoop},
  \citenamefont {Straube}, \citenamefont {Johansen},\ and\ \citenamefont
  {Tierno}}]{Stoop/etal:2020}%
  \BibitemOpen
  \bibfield  {author} {\bibinfo {author} {\bibfnamefont {R.~L.}\ \bibnamefont
  {Stoop}}, \bibinfo {author} {\bibfnamefont {A.~V.}\ \bibnamefont {Straube}},
  \bibinfo {author} {\bibfnamefont {T.~H.}\ \bibnamefont {Johansen}},\ and\
  \bibinfo {author} {\bibfnamefont {P.}~\bibnamefont {Tierno}},\ }\bibfield
  {title} {\bibinfo {title} {Collective directional locking of colloidal
  monolayers on a periodic substrate},\ }\href
  {https://doi.org/10.1103/PhysRevLett.124.058002} {\bibfield  {journal}
  {\bibinfo  {journal} {Phys. Rev. Lett.}\ }\textbf {\bibinfo {volume} {124}},\
  \bibinfo {pages} {058002} (\bibinfo {year} {2020})}\BibitemShut {NoStop}%
\bibitem [{\citenamefont {{Mirzaee-Kakhki}}\ \emph {et~al.}(2020)\citenamefont
  {{Mirzaee-Kakhki}}, \citenamefont {Ernst}, \citenamefont {{de las Heras}},
  \citenamefont {Urbaniak}, \citenamefont {Stobiecki}, \citenamefont {Tomita},
  \citenamefont {Huhnstock}, \citenamefont {Koch}, \citenamefont {G{\"o}rdes},
  \citenamefont {Ehresmann}, \citenamefont {Holzinger}, \citenamefont
  {Reginka},\ and\ \citenamefont {Fischer}}]{Mirzaee-Kakhki/etal:2020}%
  \BibitemOpen
  \bibfield  {author} {\bibinfo {author} {\bibfnamefont {M.}~\bibnamefont
  {{Mirzaee-Kakhki}}}, \bibinfo {author} {\bibfnamefont {A.}~\bibnamefont
  {Ernst}}, \bibinfo {author} {\bibfnamefont {D.}~\bibnamefont {{de las
  Heras}}}, \bibinfo {author} {\bibfnamefont {M.}~\bibnamefont {Urbaniak}},
  \bibinfo {author} {\bibfnamefont {F.}~\bibnamefont {Stobiecki}}, \bibinfo
  {author} {\bibfnamefont {A.}~\bibnamefont {Tomita}}, \bibinfo {author}
  {\bibfnamefont {R.}~\bibnamefont {Huhnstock}}, \bibinfo {author}
  {\bibfnamefont {I.}~\bibnamefont {Koch}}, \bibinfo {author} {\bibfnamefont
  {J.}~\bibnamefont {G{\"o}rdes}}, \bibinfo {author} {\bibfnamefont
  {A.}~\bibnamefont {Ehresmann}}, \bibinfo {author} {\bibfnamefont
  {D.}~\bibnamefont {Holzinger}}, \bibinfo {author} {\bibfnamefont
  {M.}~\bibnamefont {Reginka}},\ and\ \bibinfo {author} {\bibfnamefont {T.~M.}\
  \bibnamefont {Fischer}},\ }\bibfield  {title} {\bibinfo {title} {Colloidal
  trains},\ }\href {https://doi.org/10.1039/C9SM02261A} {\bibfield  {journal}
  {\bibinfo  {journal} {Soft Matter}\ }\textbf {\bibinfo {volume} {16}},\
  \bibinfo {pages} {1594} (\bibinfo {year} {2020})}\BibitemShut {NoStop}%
\bibitem [{\citenamefont {Lips}\ \emph {et~al.}(2021)\citenamefont {Lips},
  \citenamefont {Stoop}, \citenamefont {Maass},\ and\ \citenamefont
  {Tierno}}]{Lips/etal:2021}%
  \BibitemOpen
  \bibfield  {author} {\bibinfo {author} {\bibfnamefont {D.}~\bibnamefont
  {Lips}}, \bibinfo {author} {\bibfnamefont {R.~L.}\ \bibnamefont {Stoop}},
  \bibinfo {author} {\bibfnamefont {P.}~\bibnamefont {Maass}},\ and\ \bibinfo
  {author} {\bibfnamefont {P.}~\bibnamefont {Tierno}},\ }\bibfield  {title}
  {\bibinfo {title} {Emergent colloidal currents across ordered and disordered
  landscapes},\ }\href {https://doi.org/10.1038/s42005-021-00722-0} {\bibfield
  {journal} {\bibinfo  {journal} {Commun. Phys.}\ }\textbf {\bibinfo {volume}
  {4}},\ \bibinfo {pages} {224} (\bibinfo {year} {2021})}\BibitemShut {NoStop}%
\bibitem [{\citenamefont {Leyva}\ \emph {et~al.}(2022)\citenamefont {Leyva},
  \citenamefont {Stoop}, \citenamefont {Pagonabarraga},\ and\ \citenamefont
  {Tierno}}]{Leyva/etal:2022}%
  \BibitemOpen
  \bibfield  {author} {\bibinfo {author} {\bibfnamefont {S.~G.}\ \bibnamefont
  {Leyva}}, \bibinfo {author} {\bibfnamefont {R.~L.}\ \bibnamefont {Stoop}},
  \bibinfo {author} {\bibfnamefont {I.}~\bibnamefont {Pagonabarraga}},\ and\
  \bibinfo {author} {\bibfnamefont {P.}~\bibnamefont {Tierno}},\ }\bibfield
  {title} {\bibinfo {title} {Hydrodynamic synchronization and clustering in
  ratcheting colloidal matter},\ }\href
  {https://doi.org/10.1126/sciadv.abo4546} {\bibfield  {journal} {\bibinfo
  {journal} {Sci. Adv.}\ }\textbf {\bibinfo {volume} {8}},\ \bibinfo {pages}
  {eabo4546} (\bibinfo {year} {2022})}\BibitemShut {NoStop}%
\bibitem [{\citenamefont {Lips}\ \emph {et~al.}(2019)\citenamefont {Lips},
  \citenamefont {Ryabov},\ and\ \citenamefont {Maass}}]{Lips/etal:2019}%
  \BibitemOpen
  \bibfield  {author} {\bibinfo {author} {\bibfnamefont {D.}~\bibnamefont
  {Lips}}, \bibinfo {author} {\bibfnamefont {A.}~\bibnamefont {Ryabov}},\ and\
  \bibinfo {author} {\bibfnamefont {P.}~\bibnamefont {Maass}},\ }\bibfield
  {title} {\bibinfo {title} {Single-file transport in periodic potentials: {The
  Brownian} asymmetric simple exclusion process},\ }\href
  {https://doi.org/10.1103/PhysRevE.100.052121} {\bibfield  {journal} {\bibinfo
   {journal} {Phys. Rev. E}\ }\textbf {\bibinfo {volume} {100}},\ \bibinfo
  {pages} {052121} (\bibinfo {year} {2019})}\BibitemShut {NoStop}%
\bibitem [{\citenamefont {Casta{\~n}eda-Priego}\ \emph
  {et~al.}(2025)\citenamefont {Casta{\~n}eda-Priego}, \citenamefont
  {Sarmiento-G{\'o}mez}, \citenamefont {Satalsari}, \citenamefont {Egelhaaf},\
  and\ \citenamefont {Escobedo-S{\'a}nchez}}]{Castaneda-Priego/etal:2025}%
  \BibitemOpen
  \bibfield  {author} {\bibinfo {author} {\bibfnamefont {R.}~\bibnamefont
  {Casta{\~n}eda-Priego}}, \bibinfo {author} {\bibfnamefont {E.}~\bibnamefont
  {Sarmiento-G{\'o}mez}}, \bibinfo {author} {\bibfnamefont {Y.~M.}\
  \bibnamefont {Satalsari}}, \bibinfo {author} {\bibfnamefont {S.~U.}\
  \bibnamefont {Egelhaaf}},\ and\ \bibinfo {author} {\bibfnamefont {M.~A.}\
  \bibnamefont {Escobedo-S{\'a}nchez}},\ }\bibfield  {title} {\bibinfo {title}
  {Colloidal transport in periodic potentials: the role of
  modulated-crowding},\ }\href {https://doi.org/10.1039/D5SM00133A} {\bibfield
  {journal} {\bibinfo  {journal} {Soft Matter}\ }\textbf {\bibinfo {volume}
  {21}},\ \bibinfo {pages} {3868} (\bibinfo {year} {2025})}\BibitemShut
  {NoStop}%
\bibitem [{\citenamefont {Lips}\ \emph {et~al.}(2018)\citenamefont {Lips},
  \citenamefont {Ryabov},\ and\ \citenamefont {Maass}}]{Lips/etal:2018}%
  \BibitemOpen
  \bibfield  {author} {\bibinfo {author} {\bibfnamefont {D.}~\bibnamefont
  {Lips}}, \bibinfo {author} {\bibfnamefont {A.}~\bibnamefont {Ryabov}},\ and\
  \bibinfo {author} {\bibfnamefont {P.}~\bibnamefont {Maass}},\ }\bibfield
  {title} {\bibinfo {title} {Brownian asymmetric simple exclusion process},\
  }\href {https://doi.org/10.1103/PhysRevLett.121.160601} {\bibfield  {journal}
  {\bibinfo  {journal} {Phys. Rev. Lett.}\ }\textbf {\bibinfo {volume} {121}},\
  \bibinfo {pages} {160601} (\bibinfo {year} {2018})}\BibitemShut {NoStop}%
\bibitem [{\citenamefont {Derrida}(1998)}]{Derrida:1998}%
  \BibitemOpen
  \bibfield  {author} {\bibinfo {author} {\bibfnamefont {B.}~\bibnamefont
  {Derrida}},\ }\bibfield  {title} {\bibinfo {title} {An exactly soluble
  non-equilibrium system: The asymmetric simple exclusion process},\ }\href
  {https://doi.org/https://doi.org/10.1016/S0370-1573(98)00006-4} {\bibfield
  {journal} {\bibinfo  {journal} {Phys. Rep.}\ }\textbf {\bibinfo {volume}
  {301}},\ \bibinfo {pages} {65 } (\bibinfo {year} {1998})}\BibitemShut
  {NoStop}%
\bibitem [{\citenamefont {Sch\"utz}(2001)}]{Schuetz:2001}%
  \BibitemOpen
  \bibfield  {author} {\bibinfo {author} {\bibfnamefont {G.~M.}\ \bibnamefont
  {Sch\"utz}},\ }\bibfield  {title} {\bibinfo {title} {Exactly solvable models
  for many-body systems far from equilibrium},\ }in\ \href
  {https://doi.org/https://doi.org/10.1016/S1062-7901(01)80015-X} {\emph
  {\bibinfo {booktitle} {Phase Transitions and Critical Phenomena}}},\
  Vol.~\bibinfo {volume} {19},\ \bibinfo {editor} {edited by\ \bibinfo {editor}
  {\bibfnamefont {C.}~\bibnamefont {Domb}}\ and\ \bibinfo {editor}
  {\bibfnamefont {J.}~\bibnamefont {Lebowitz}}}\ (\bibinfo  {publisher}
  {Academic Press},\ \bibinfo {address} {London},\ \bibinfo {year} {2001})\
  pp.\ \bibinfo {pages} {1--251}\BibitemShut {NoStop}%
\bibitem [{\citenamefont {Schmittmann}\ and\ \citenamefont
  {Zia}(1998)}]{Schmittmann/Zia:1998}%
  \BibitemOpen
  \bibfield  {author} {\bibinfo {author} {\bibfnamefont {B.}~\bibnamefont
  {Schmittmann}}\ and\ \bibinfo {author} {\bibfnamefont {R.~K.~P.}\
  \bibnamefont {Zia}},\ }\bibfield  {title} {\bibinfo {title} {Driven diffusive
  systems. {A}n introduction and recent developments},\ }\href
  {https://doi.org/https://doi.org/10.1016/S0370-1573(98)00005-2} {\bibfield
  {journal} {\bibinfo  {journal} {Phys. Rep.}\ }\textbf {\bibinfo {volume}
  {301}},\ \bibinfo {pages} {45} (\bibinfo {year} {1998})}\BibitemShut
  {NoStop}%
\bibitem [{\citenamefont {Mallick}(2015)}]{Mallick:2015}%
  \BibitemOpen
  \bibfield  {author} {\bibinfo {author} {\bibfnamefont {K.}~\bibnamefont
  {Mallick}},\ }\bibfield  {title} {\bibinfo {title} {The exclusion process: A
  paradigm for non-equilibrium behaviour},\ }\href
  {https://doi.org/https://doi.org/10.1016/j.physa.2014.07.046} {\bibfield
  {journal} {\bibinfo  {journal} {Physica A}\ }\textbf {\bibinfo {volume}
  {418}},\ \bibinfo {pages} {17} (\bibinfo {year} {2015})},\ \bibinfo {note}
  {proceedings of the 13th International Summer School on Fundamental Problems
  in Statistical Physics}\BibitemShut {NoStop}%
\bibitem [{\citenamefont {Antonov}\ \emph {et~al.}(2022)\citenamefont
  {Antonov}, \citenamefont {Ryabov},\ and\ \citenamefont
  {Maass}}]{Antonov/etal:2022a}%
  \BibitemOpen
  \bibfield  {author} {\bibinfo {author} {\bibfnamefont {A.~P.}\ \bibnamefont
  {Antonov}}, \bibinfo {author} {\bibfnamefont {A.}~\bibnamefont {Ryabov}},\
  and\ \bibinfo {author} {\bibfnamefont {P.}~\bibnamefont {Maass}},\ }\bibfield
   {title} {\bibinfo {title} {Solitons in overdamped {B}rownian dynamics},\
  }\href {https://doi.org/10.1103/PhysRevLett.129.080601} {\bibfield  {journal}
  {\bibinfo  {journal} {Phys. Rev. Lett.}\ }\textbf {\bibinfo {volume} {129}},\
  \bibinfo {pages} {080601} (\bibinfo {year} {2022})}\BibitemShut {NoStop}%
\bibitem [{\citenamefont {Cereceda-L\'opez}\ \emph {et~al.}(2023)\citenamefont
  {Cereceda-L\'opez}, \citenamefont {Antonov}, \citenamefont {Ryabov},
  \citenamefont {Maass},\ and\ \citenamefont
  {Tierno}}]{Cereceda-Lopez/etal:2023}%
  \BibitemOpen
  \bibfield  {author} {\bibinfo {author} {\bibfnamefont {E.}~\bibnamefont
  {Cereceda-L\'opez}}, \bibinfo {author} {\bibfnamefont {A.~P.}\ \bibnamefont
  {Antonov}}, \bibinfo {author} {\bibfnamefont {A.}~\bibnamefont {Ryabov}},
  \bibinfo {author} {\bibfnamefont {P.}~\bibnamefont {Maass}},\ and\ \bibinfo
  {author} {\bibfnamefont {P.}~\bibnamefont {Tierno}},\ }\bibfield  {title}
  {\bibinfo {title} {Overcrowding induces fast colloidal solitons in a slowly
  rotating potential landscape},\ }\href
  {https://doi.org/10.1038/s41467-023-41989-x} {\bibfield  {journal} {\bibinfo
  {journal} {Nat. Commun.}\ }\textbf {\bibinfo {volume} {14}},\ \bibinfo
  {pages} {6448} (\bibinfo {year} {2023})}\BibitemShut {NoStop}%
\bibitem [{\citenamefont {Mishra}\ \emph {et~al.}(2025)\citenamefont {Mishra},
  \citenamefont {Ryabov},\ and\ \citenamefont {Maass}}]{Mishra/etal:2025}%
  \BibitemOpen
  \bibfield  {author} {\bibinfo {author} {\bibfnamefont {S.}~\bibnamefont
  {Mishra}}, \bibinfo {author} {\bibfnamefont {A.}~\bibnamefont {Ryabov}},\
  and\ \bibinfo {author} {\bibfnamefont {P.}~\bibnamefont {Maass}},\ }\bibfield
   {title} {\bibinfo {title} {Phase locking and fractional shapiro steps in
  collective dynamics of microparticles},\ }\href
  {https://doi.org/10.1103/PhysRevLett.134.107102} {\bibfield  {journal}
  {\bibinfo  {journal} {Phys. Rev. Lett.}\ }\textbf {\bibinfo {volume} {134}},\
  \bibinfo {pages} {107102} (\bibinfo {year} {2025})}\BibitemShut {NoStop}%
\bibitem [{\citenamefont {Antonov}\ \emph {et~al.}(2024)\citenamefont
  {Antonov}, \citenamefont {Ryabov},\ and\ \citenamefont
  {Maass}}]{Antonov/etal:2024}%
  \BibitemOpen
  \bibfield  {author} {\bibinfo {author} {\bibfnamefont {A.~P.}\ \bibnamefont
  {Antonov}}, \bibinfo {author} {\bibfnamefont {A.}~\bibnamefont {Ryabov}},\
  and\ \bibinfo {author} {\bibfnamefont {P.}~\bibnamefont {Maass}},\ }\bibfield
   {title} {\bibinfo {title} {Solitary cluster waves in periodic potentials:
  Formation, propagation, and soliton-mediated particle transport},\ }\href
  {https://doi.org/https://doi.org/10.1016/j.chaos.2024.115079} {\bibfield
  {journal} {\bibinfo  {journal} {Chaos, Solitons \& Fractals}\ }\textbf
  {\bibinfo {volume} {185}},\ \bibinfo {pages} {115079} (\bibinfo {year}
  {2024})}\BibitemShut {NoStop}%
\bibitem [{\citenamefont {Antonov}\ \emph {et~al.}(2025)\citenamefont
  {Antonov}, \citenamefont {Schweers}, \citenamefont {Ryabov},\ and\
  \citenamefont {Maass}}]{Antonov/etal:2025}%
  \BibitemOpen
  \bibfield  {author} {\bibinfo {author} {\bibfnamefont {A.~P.}\ \bibnamefont
  {Antonov}}, \bibinfo {author} {\bibfnamefont {S.}~\bibnamefont {Schweers}},
  \bibinfo {author} {\bibfnamefont {A.}~\bibnamefont {Ryabov}},\ and\ \bibinfo
  {author} {\bibfnamefont {P.}~\bibnamefont {Maass}},\ }\bibfield  {title}
  {\bibinfo {title} {Fast {B}rownian cluster dynamics},\ }\href
  {https://doi.org/https://doi.org/10.1016/j.cpc.2024.109474} {\bibfield
  {journal} {\bibinfo  {journal} {Comput. Phys. Commun.}\ }\textbf {\bibinfo
  {volume} {309}},\ \bibinfo {pages} {109474} (\bibinfo {year}
  {2025})}\BibitemShut {NoStop}%
\bibitem [{\citenamefont {Abramowitz}\ and\ \citenamefont
  {Stegun}(1965)}]{Abramowitz/Stegun:1965}%
  \BibitemOpen
  \bibfield  {author} {\bibinfo {author} {\bibfnamefont {M.}~\bibnamefont
  {Abramowitz}}\ and\ \bibinfo {author} {\bibfnamefont {I.}~\bibnamefont
  {Stegun}},\ }\href {https://books.google.de/books?id=MtU8uP7XMvoC} {\emph
  {\bibinfo {title} {Handbook of Mathematical Functions: With Formulas, Graphs,
  and Mathematical Tables}}},\ Applied mathematics series\ (\bibinfo
  {publisher} {Dover Publications},\ \bibinfo {year} {1965})\BibitemShut
  {NoStop}%
\bibitem [{Note1()}]{Note1}%
  \BibitemOpen
  \bibinfo {note} {There exists also a variant of the soliton propagation,
  where during the motion of the composite cluster an $n_{\protect \rm
  b}$-cluster first attaches at its back end and shortly after detaches \cite
  {Antonov/etal:2024}.}\BibitemShut {Stop}%
\bibitem [{Note2()}]{Note2}%
  \BibitemOpen
  \bibinfo {note} {In Ref.~\cite {Antonov/etal:2024}, there is $n_{\protect \rm
  b}/\protect \qopname \relax m{gcd}(n_{\protect \rm b},n_{\protect \rm c})$ in
  the denominator of Eq.~\protect \textup {\hbox {\mathsurround \z@ \protect
  \normalfont (\ignorespaces \ref {eq:Jformer}\unskip \@@italiccorr )}} instead
  of $n_{\protect \rm b}$. However, it was shown that $n_{\protect \rm b}$ and
  $n_{\protect \rm b}+n_{\protect \rm c}$ are coprime and hence $n_{\protect
  \rm b}$ and $n_{\protect \rm c}$ are coprime also, i.e.\ their greatest
  common divisor satisfies $\protect \qopname \relax m{gcd}(n_{\protect \rm
  b},n_{\protect \rm c})=\protect \qopname \relax m{gcd}(n_{\protect \rm
  b}+n_{\protect \rm c},n_{\protect \rm c})=1$.}\BibitemShut {Stop}%
\bibitem [{\citenamefont {Vorobyov}(1980)}]{Vorobyov:1980}%
  \BibitemOpen
  \bibfield  {author} {\bibinfo {author} {\bibfnamefont {N.~N.}\ \bibnamefont
  {Vorobyov}},\ }\href@noop {} {\emph {\bibinfo {title} {Criteria for
  divisibility}}}\ (\bibinfo  {publisher} {University of Chicago Press},\
  \bibinfo {year} {1980})\BibitemShut {NoStop}%
\end{thebibliography}

%

\end{document}